# Quantum Fluctuation-enhanced Milli-Kelvin Magnetic Refrigeration in Triangular Lattice Magnet GdBO$_3$

Weijie Lin[1,3,5,‡], Nan Zhao[1,4,6,‡], Zhaoyi Li[1,3,5], Weiran An[1,3], Ruixin Guo[1,3], Jianqiao Wang[1,3], Changzhao Pan[3], Bo Wen[5], Jieming Sheng[2,*], Liusuo Wu[1,*], Shu Guo[1,3,*]

[1]Shenzhen Institute for Quantum Science and Engineering, Department of Chemistry, and Department of Physics, Southern University of Science and Technology, Shenzhen 518055, China.

[2]School of Physical Sciences, Great Bay University and Great Bay Institute for Advanced Study, Dongguan,523000, China.

[3]International Quantum Academy, Shenzhen 518048, China.

[4]Institute of High Energy Physics, Chinese Academy of Sciences (CAS), Beijing 100049, China.

[5]School of Physics and Electronics, Henan University, Kaifeng 475004, China.

[6]Spallation Neutron Source Science Center, Dongguan 523803, China

[‡]These authors contributed equally to this work.
[*]Correspondence should be addressed to J. Sheng (shengjm@gbu.edu.cn), L. Wu (wuls@sustech.edu.cn), and S. Guo (guos@sustech.edu.cn).

Keywords: geometrical frustration, magnetic refrigeration, triangular lattice, rare earth

Rare-earth-based triangular lattice antiferromagnets, with strong quantum fluctuations and weak magnetic interactions, can often retain large magnetic entropy down to very low temperatures, making them excellent candidates for magnetic refrigeration at ultra-low temperatures. These materials exhibit a substantial magnetocaloric effect (MCE) due to enhanced spin fluctuations, particularly near quantum critical points, which leads to significant changes in magnetic entropy. This study reports on the crystal growth, structure, magnetism, and MCE of a Gd-based triangular lattice material, $GdBO_3$, characterized by a large spin quantum number ($S = 7/2$). Successive phase transitions ($T_1 = 0.52$ K, $T_2 = 0.88$ K, and $T_3 = 1.77$ K) were observed in zero-field specific heat measurements. Furthermore, thermal dynamic analysis under external magnetic fields identified five distinct phase regions and three quantum critical points for $GdBO_3$. Due to its broad specific heat features and the high density of magnetic $Gd^{3+}$ ions, we achieved a minimum temperature of 50 mK near the field-induced quantum critical point, using a custom-designed $GdBO_3$-based adiabatic demagnetization refrigerator. Our findings reveal significant quantum fluctuations below 2 K, demonstrating $GdBO_3$'s potential for milli-Kelvin magnetic cooling applications.

## 1. Introduction

The magnetocaloric effect (MCE) at ultra-low temperatures is of significant interest for both fundamental research,[1] and practical applications,[2] particularly in low-temperature physics, refrigeration, and quantum technologies.[3] In triangular lattice (TL) antiferromagnetic (AFM) systems, competing interactions between neighboring spins often lead to strong quantum fluctuations and the preservation of substantial magnetic entropy at low temperatures.[4] Through adiabatic demagnetization, especially near quantum critical points, these systems can exhibit a pronounced MCE at ultra-low temperatures.[5] Recently, Gang Su et al. successfully established milli-Kelvin temperatures via adiabatic demagnetization refrigeration (ADR) down to 94 mK by leveraging the MCE of the spin supersolid state in the $Co^{2+}$-based TL magnet $Na_2BaCo(PO_4)_2$ ($S = 1/2$).[1b] Unlike an $S = 1/2$ system, the commercial garnet $Gd_3Ga_5O_{13}$ with $S = 7/2$ is widely used for low-temperature ADR applications with a high density of magnetic ions and strong cooling powder.[6] It could be considered a three-dimensional hyper-kagome lattice with the hidden order in a spin liquid state.[7] In classical water-containing paramagnetic salts like $Ce_2Mg_3(NO_3)_{12} \cdot 24H_2O$ (CMN) and $Fe(SO_4)_2(NH_4) \cdot 12H_2O$ (FAA),[8] the large distances between magnetic ions are mediated by bridging water molecules, which weaken the magnetic interactions and lower the magnetic transition to the milli-Kelvin range. However, the cooling capacity of CMN and FAA is constrained by the low density of magnetic ions.

Moreover, dehydration of CMN and FAA in vacuum or mild heating hinders their application. In contrast, the geometrically frustrated system (GFM) offers a superior balance between a low ADR limit temperature and a high density of magnetic ions. Inspired by this, $Gd^{3+}$-based GFMs, such as TLs with a larger quantum spin number ($S = 7/2$), may exhibit enhanced MCE at ultra-low temperatures, making them promising candidates for advanced refrigeration technologies.

The magnetic properties of $RE$BO$_3$ ($RE$ = Eu-Yb) materials with rare earth ($RE$)-based TL have recently attracted widespread attention.[9] In particular, GdBO$_3$ exhibits multi-level phase transitions under zero field, which indicates strong quantum fluctuations.[9a] The large magnetic entropy difference, $\Delta S_m$ (2 K, 9 T→0 T) = 57.8 J mol$^{-1}$ K$^{-1}$, of GdBO$_3$ calculated by Maxwell relation based on polycrystalline samples, preliminarily confirmed that GdBO$_3$ may have certain prospects for magnetic refrigeration.[9a] Subsequently, Shen Jun et al. increased the $\Delta S_m$ (> 2 K) of GdBO$_3$ under low fields by doping with $Dy^{3+}$ ions.[9e] The above evidence preliminarily reveals interesting quantum magnetism and promising milli-Kelvin ADR applications for GdBO$_3$. However, the lack of single crystals and systematic magnetic properties characterizations has brought difficulties to further physical understanding and application for GdBO$_3$.

In this work, the anisotropic magnetism, specific heat, and MCE of GdBO$_3$ were studied through single crystals. The substantial Curie-Weiss (C-W) temperature ($\Theta_\perp$ = -7.0 K) derived from magnetic susceptibility analysis suggests relatively weak antiferromagnetic interactions among $Gd^{3+}$ cations for GdBO$_3$. Additionally, the systematic phase diagram of GdBO$_3$ was established through the combination of temperature- and field-dependent specific heat measurements. The magnetic states and MCE of each region were explained through magnetic entropy analysis. More importantly, we discovered a pronounced quantum critical enhanced MCE in GdBO$_3$, which enabled us to achieve a minimum temperature of 50 mK using a custom-designed GdBO$_3$-based ADR device.

## 2. Results and Discussion
## 2.1. Crystal structure of GdBO$_3$

Millimeter-size single crystals of GdBO$_3$ (**Figure 1a**) were grown using the high-temperature flux method and subsequently used for structural and magnetic property characterizations. It crystallized in a monoclinic crystal system with a space group of $C2/c$ (no. 15), which is similar to the previous results of $RE$BO$_3$ ($RE$ = Tb-Yb)[10] from Single Crystal X-ray diffraction (SC-XRD), as shown in **Figure 1b**. In the magnetic picture, two unique positions are found for Gd ions (Gd1 and Gd2), which are further arranged in the TL plane (**Figure 1c**). The experimental Powder X-ray diffraction (P-XRD) patterns are in good agreement with the Rietveld refinements profiles through the GSAS-II (**Figure 1d**). The magnetic framework is constructed from GdO$_8$ polyhedra, connected through edge-sharing within the TL layers and corner-sharing between adjacent TL layers. Consequently, the distance between Gd ions within a single TL layer (denoted as $d_{intra}$(Gd-Gd)) ranges from 3.79 Å to 3.88 Å. In contrast, the distance between Gd ions across different layers (denoted as $d_{inter}$(Gd-Gd)) extends slightly further, measuring between 4.45 Å and 4.46 Å. The detailed structural information is summarized in **Tables S1** and **S2**. As reported by Zhi-Jun Zhang et al.,[11] we also found a small amount of oxygen defects based on IR spectra (**Figure S1**).

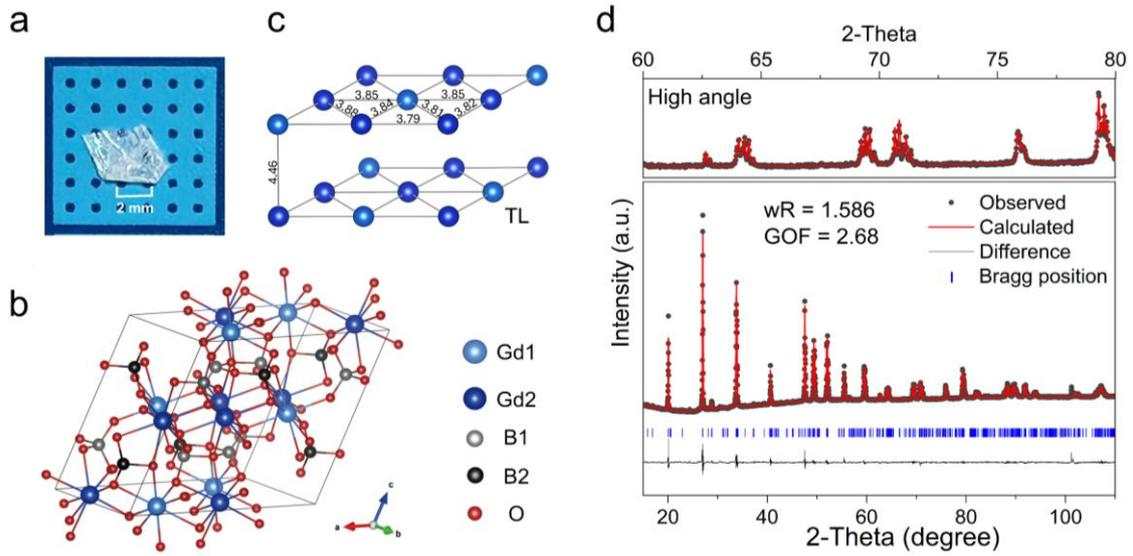

**Figure 1 Crystal features, Crystal structure, and Phase Purity of GdBO$_3$.** (a) Crystal picture of GdBO$_3$. (b) Unit cell and (c) magnetic framework of GdBO$_3$ with the nearest inter-plane distance ($d_{inter}$ = 4.45 - 4.46 Å) larger than that in the TL plane ($d_{intra}$ = 3.79-3.88 Å). (d) Rietveld refinement plot of P-XRD patterns and simulated patterns from the SC-XRD results at 100 K. The up-inset shows the comparison of experimental P-XRD (black dots).

## 2.2. Magnetism of GdBO$_3$

Considering the different magnetic exchange paths within and between the TL layers, anisotropic magnetic measurements were performed with the magnetic field $B$ perpendicular ($B_\perp$) and parallel ($B_\parallel$) to the TL layers. As shown in **Figure 2a**, $\chi^{-1}(T)$ varies linearly with temperature, indicating well C-W paramagnetism for GdBO$_3$ under an external magnetic field of 0.2 T. In addition, no signs of magnetic order were observed above 2 K. Therefore, the C-W paramagnetic model, given by $\chi = \frac{C}{T-\Theta} + \chi_0$ (where Θ is the C-W temperature of the material and $\chi_0$ represents the temperature-independent magnetic susceptibility contribution), is used to fit the magnetic susceptibility of GdBO$_3$ for both directions over the temperature range of 10-40 K. The effective magnetic moment, $\mu_{eff}$, is calculated using the equation $\mu_{eff} = \sqrt{8C}\mu_B$. The $\mu_{eff}$ calculated based on the magnetic fields $B_\parallel$ and $B_\perp$ are 8.0 $\mu_B$ and 8.2 $\mu_B$, respectively, which is consistent with the Gd$^{3+}$ (4$f^{\,7}$) magnetic moment $\mu_{cal} = 7.9$ $\mu_B$ predicted by Hund's rules. The substantial values of C-W temperatures ($\Theta_\parallel = -4.8$ K, $\Theta_\perp = -7.0$ K) suggests relatively weak antiferromagnetic exchange interactions among Gd$^{3+}$ ions.

Field-dependent magnetization measurements were performed on GdBO$_3$ with magnetic fields $B_{//}$ and $B_\perp$ at 2 K based on the same single crystal, as shown in **Figure 2b** and **Figure S2**. The saturated moment in both directions is nearly at 6.99 $\mu_B$, which is close to the theoretically predicted saturation magnetic moment of 7.0 $\mu_B$ of free Gd$^{3+}$ ions ($L = 0$, $S = 7/2$). Compared to the magnetic field applied along the *c*-axis, an in-plane magnetic field more easily polarizes the spins, suggesting a weak easy-plane magnetic anisotropy in GdBO$_3$, a behavior also observed in KBaGd(BO$_3$)$_2$.[12] It is generally assumed that the total orbital angular momentum is zero for Gd-based compounds, leading to isotropic magnetic behavior. However, recent studies have shown that in Gd$^{3+}$ ions, the contribution of high-energy mixed states with $L \neq 0$ cannot be neglected, even at the energy level of 1-2 K.[13] Additional experiments, such as inelastic neutron scattering and electron spin resonance, are essential to better understand the origin of this anisotropy and clarify the slightly easy-plane magnetic behavior observed in GdBO$_3$.

Isothermal field-dependent magnetizations were conducted in the temperature range of 1.8–2.2 K. An anomaly, shown in **Figure S3**, reveals that the magnetization at 1.8 K is lower than at 1.9 K for $B_\perp = 1.6$ T. To understand this anomaly, **Figure 2c** shows the change of magnetic susceptibilities with magnetic fields at different temperatures. When the temperature is $T = 2.2$ K, the magnetic susceptibility decreases gradually, which is typical paramagnetic behavior (**I**). However, when the temperature drops below $T = 2$ K, the magnetic susceptibility decreases sharply at field $B_\perp = 1.6$ T. Moreover, $B_\perp = 1.5$ T was applied for temperature-dependent magnetic susceptibility, as shown in **Figure 2d**. As the temperature gradually decreases, the magnetic susceptibility decreases significantly at $T = 1.92$ K. Corresponding to the two negative C-W temperatures ($\Theta_\parallel = -4.8$ K and $\Theta_\perp = -7.0$ K), GdBO$_3$ is experiencing the antiferromagnetic spin pairing transition at $B_\perp = 1.6$ T, $T = 1.8$ K.

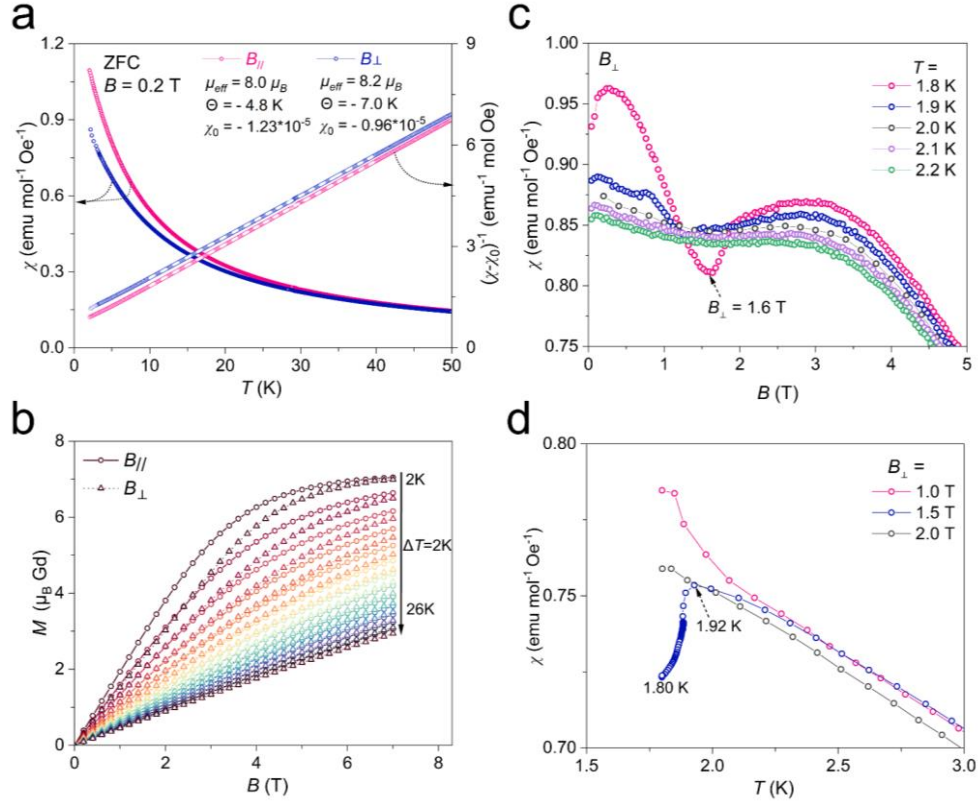

**Figure 2 Anisotropic Magnetism of GdBO$_3$.** (a) Temperature-dependent DC magnetic susceptibilities measured by PPMS for $B_{\parallel}$ and $B_{\perp}$, respectively. (b) Field-dependence of magnetization at different temperatures for $B_{\parallel}$ (solid lines) and $B_{\perp}$ (dash lines) measured by MPMS3. (c) Field-dependent magnetic susceptibilities with $B_{\perp}$. (d) Temperature-dependent magnetic susceptibility performed at 0 T, 1.5 T, and 2 T.

## 2.3. Specific heat of GdBO$_3$

The specific heat of GdBO$_3$ with the magnetic field ($B_{\perp}$ = 0-1.5 T) and of the non-magnetic isostructural compound LuBO$_3$ are shown in **Figure 3a**. When the temperature is lower than 5 K, the specific heat value of LuBO$_3$ is close to zero. Hence, the phonons contribution of GdBO$_3$ could be overlooked below 5 K. Since GdBO$_3$ is a transparent insulator, its electron specific heat is also negligible.[14] Therefore, it can be inferred that the specific heat $C_p$ measured below 5 K is predominantly due to magnetic contributions, with $C_{mag} \approx C_p$. The inset of **Figure 3a** reveals the presence of a sharp $\lambda$-type peak at $T_{N3}$ = 1.77 K, accompanied by two anomalies at $T_{N2}$ = 0.88 K and $T_{N1}$ = 0.52 K under zero field, respectively. It is worth noting that previous zero-field specific heat measurements on polycrystalline GdBO$_3$ samples by P. Mukherjee et al. (2018) [9a] only identified two transitions at 1.72 K and 0.61 K, respectively. These discrepancies may arise from averaging effects inherent to polycrystalline samples. Interestingly, as the field increases from 0 T to 1 T (see **Figure 3a**), the $T_{N3}$ shifts from 1.77 K to 1.87 K and the peak gradually becomes higher and sharper. Meanwhile, the transitions at $T_{N1}$ and $T_{N2}$ are gradually suppressed, with a single characteristic peak observed at $B_\perp = 0.5$ T. By 1 T, only the $T_{N3}$ peak remains. As the magnetic field increases further (1.5 T to 3 T), $T_{N3}$ approaches 0 K, as shown in **Figure 3b**, indicating an antiferromagnetic transition.

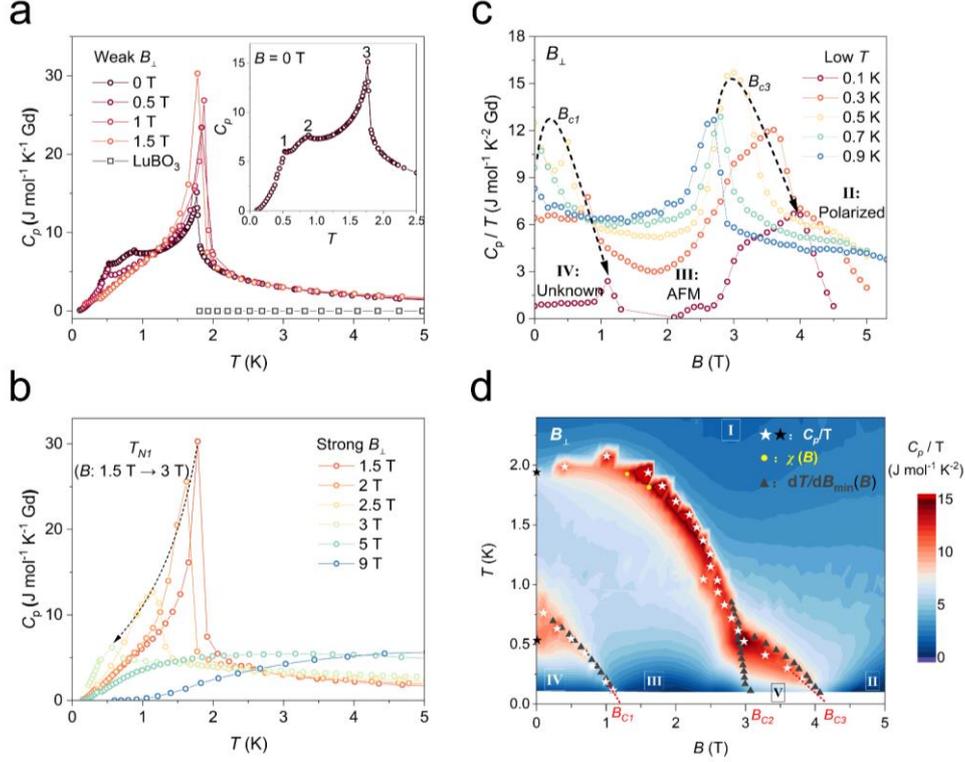

**Figure 3. Thermal Dynamical Properties and Magnetic Phase Diagram of GdBO$_3$.** Temperature-dependence specific heat of GdBO$_3$ measured at (a) weak fields, and (b) strong fields. (c) Field-dependence specific heat measured under 0.1–0.9 K. (d) The magnetic phase diagram of GdBO$_3$.

Moreover, field-dependent specific heat measurements of GdBO$_3$ were conducted using the dilution refrigerator (DR) of the PPMS system at various temperatures, as shown in **Figures 3c** and **S4**. As the temperature decreases from 0.9 K to 0.1 K, the critical field $B_{c3}$ gradually increases (**Figure 3c**). At $T = 0.1$ K, the value of $C_p/T$ is very close to zero in the range of $B_\perp = $ 1.5-2.5 T. According to previous magnetic and thermodynamic behaviors, GdBO$_3$ forms an ordered ground state with antiferromagnetic (**III**) spin alignment. The magnetic moments become polarized (**II**) when the magnetic field reaches 4.5 T. It is worth noting that within the low-field range of $B_\perp = $ 0–1 T, a distinct magnetic order (**IV**) emerges, clearly differentiated from the ground state phase (**III**) observed in the higher field range of $B_\perp = $ 1.5–2.5 T. This is attributed to the phase transitions associated with $T_{N1}$ and $T_{N2}$. In the future, elastic neutron scattering will be essential to identifying these diverse magnetic spin states, once the neutron absorption challenges of Gd and B are addressed.

The magnetic phase diagram overplotted on the contour plot of $C_p/T$ data is presented in **Figure 3d**. The anomalies observed in the $C_p(T, B)/T$ curves are highlighted with white five-pointed stars for clarity. The white stars can represent the critical point ($B_c$) and divide the phase diagram into several regions (**I - IV**). The $B_c$ separating each ground state is highly disordered. The white stars in the phase diagram are connected, and the trend is linearly extrapolated down to 0 K, as shown by the red dotted line in **Figure 3d**. The two red dotted lines intersect the lower coordinate axis (0 K) at two critical points $B_{c1}$ = 1.25 T and $B_{c3}$ = 4.3 T, respectively. More significantly, the pronounced quantum fluctuations near the quantum critical points offer strong evidence for the potential use of GdBO$_3$ in milli-Kelvin ADR applications.

## 2.4. MCE of GdBO$_3$

In an insulating material, the total entropy ($S_{tot}$) is contributed by phonons ($S_{lat}$) and magnons ($S_{mag}$), that is: $S_{tot} \approx S_{lat} + S_{mag}$. Since the phonon-specific heat of GdBO$_3$ is negligible below 5 K, the $S_{lat}$ is also equal to zero, and the total entropy is almost composed of magnetic entropy ($S_{tot} \approx S_{mag}$). The magnetic entropy of GdBO$_3$ could be evaluated using the formula $S_{mag} = \int_0^T \frac{C_{mag}}{T} dT$, as shown in **Figures 4a** and **S5**. In adiabatic conditions, since the $S_{tol}(T, B)$ is conserved, changes in the external magnetic field will generate MCE. The strength of the MCE is related to the steady-state magnetic entropy difference under initial and final magnetic fields ($B_i$ and $B_f$), that is, $\Delta S_{mag}(T, B_i, B_f) = S_{mag}(T, B_i) - S_{mag}(T, B_f)$.

As shown in **Figure S6**, under $B_i = 9T, B_f = 0T$, there is a large entropy difference from 0.1 K to 5 K. In particular, the $-\Delta S_{mag}$ value at $T$ = 2.67 K reaches 57.4 J kg$^{-1}$ K$^{-1}$, indicating a giant MCE. The magnetic entropy difference between different $B_i$ and $B_f$ = 0 T is calculated, as shown in **Figure S7**. When $B_i$ = 5 T, 7 T, 9 T, $-\Delta S_{mag}$ is a relatively large positive value. Comparing its maximum value when $B_i$ = 5 T, 7 T, 9 T with that of well-known magnetic refrigerants (**Table 1**), it is found that GdBO$_3$ exhibits a relatively large $-\Delta S_{mag}$. Interestingly, as illustrated in the magnetic entropy phase diagram (**Figure 4a**), the strong quantum fluctuations near $B_{c3}$ lead to a substantial increase in magnetic entropy. Consequently, setting the final magnetic field to 3.5 T significantly enhances the magnitude of the MCE, as demonstrated in the calculation results (**Figure 4b**). Below 0.6 K, when the initial magnetic field $B_i$ is less than or greater than 3.5 T, there is a large and positive MCE. It is worth noting that the magnetic refrigeration range of GdBO$_3$ is lower than 0.1 K due to significant residual entropy persisting below the experimental temperature limit.

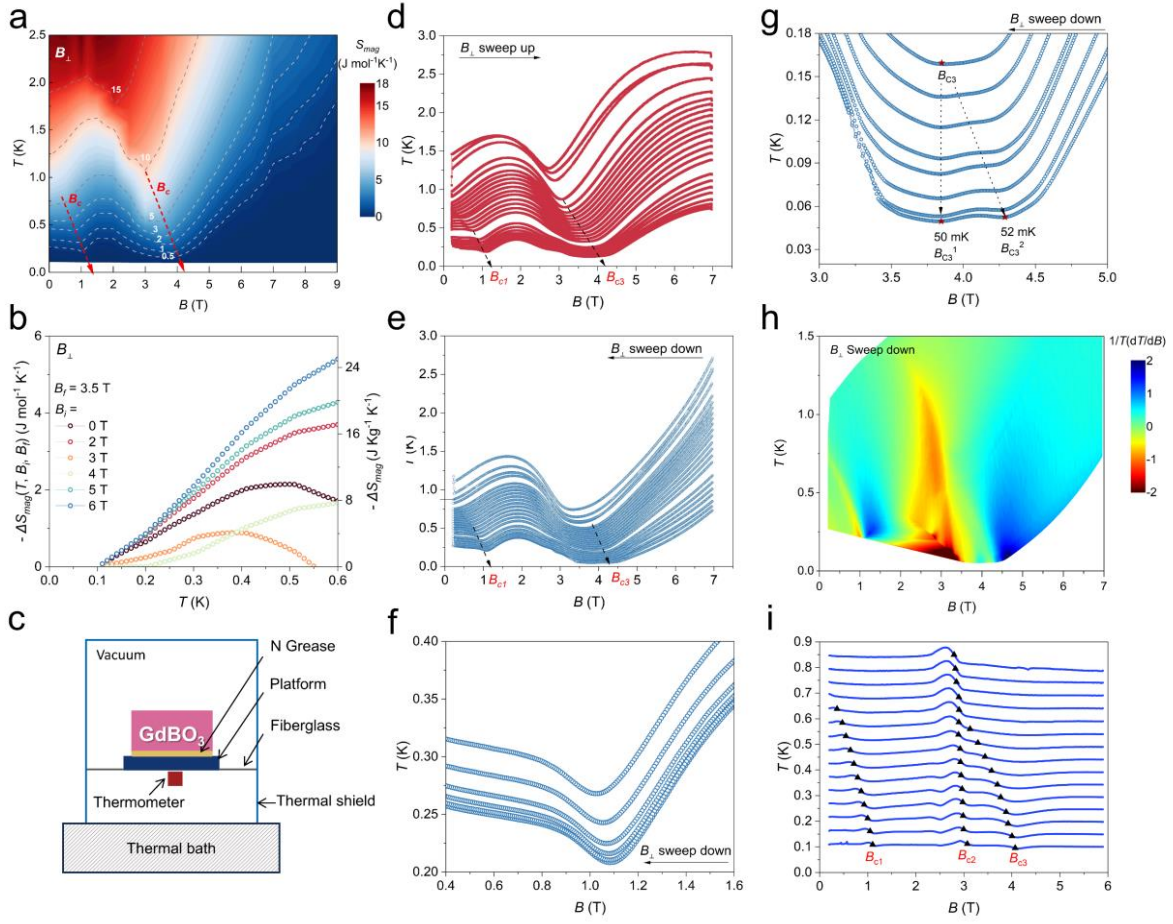

**Figure 4**. **ADR Device and Cooling Performances of GdBO$_3$ in the Commercial PPMS system**. (a) Magnetic entropy phase diagram of GdBO$_3$. The gray dashed line marks the isentropic point. (b) Magnetic entropy changes result in $\Delta S_{mag}$ vs. $T$ of GdBO$_3$ under different final fields 3.5 T. (c) ADR device. Temperature changes with (d) increasing, and (e) decreasing magnetic fields, based on 10 mg samples. Temperature changes with the magnetic field at the quantum critical point (f) $B_{c1}$ and (g) $B_{c3}$. (h) Grüneisen parameters of GdBO$_3$ at different temperatures and magnetic fields. (i) MCE based on 0.05 mg sample.

The MCE was measured based on single crystals of GdBO$_3$ with a mass of 10 mg under quasi-adiabatic conditions using a self-designed ADR device (**Figure 4c**). A field rate of 60 Oe/s was used for both magnetization (**Figure 4d**) and demagnetization (**Figure 4e**), recording the temperature $T(B)$ under different magnetic fields. The minimum of the $T(B)$ all appear near the $B_{c1}$ and $B_{c3}$. The overall shape of the $T(B)$ matches well with the contours of magnetic entropy in **Figure 4a**. As shown in **Figure 4e**, the temperature changes as high as 2 K with the initial parameter ($B_i$ = 7 T and $T_i$ = 2.7 K). **Figures 4f** and **4g** show the detailed change of $T(B)$ with a magnetic field $B_{c1}$ and $B_{c3}$. As shown in **Figure 4f**, as the magnetic field gradually decreases, the $T(B)$ of GdBO$_3$ decreases rapidly until it reaches the lowest point around 1.1 T, which indicates the transition from phase **III** to phase **IV**.

Different from $B_{c1}$, a wide trough is observed near $B_{c3}$, and the minimum temperature value reaches 50 mK (**Figure 4g**). It was found that the width of the trough (3.5 - 4.5 T) is as high as 1 T, indicating a highly degenerated magnetic state. This happens because the magnetic field energy is comparable to the antiferromagnetic exchange interaction, making it difficult for the spins to establish either a polarized ground state or an antiferromagnetic state. Moreover, there is only one minimum in the curve at relatively high temperatures. However, in the lower temperature range, two minima appear, indicated by red star markers in **Figure 4g**. This implies that the $B_{c2}$ may give rise to two previously unknown phase transitions ($B_{c3}^1$ and $B_{c3}^2$), potentially revealing an unidentified state between them. Next, Grüneisen parameters, $\Gamma_B = 1/T \, (dT/dB)$, were used to describe the magnetic cooling and magnetothermal regions of GdBO$_3$. As illustrated in **Figure 4h**, a positive value (represented in blue) indicates a decrease in temperature as the magnetic field is reduced. It can be found that excellent temperature regulation can be achieved near $B_{c1}$ and $B_{c3}$.

However, due to the significant MCE from the larger crystal, it is difficult to observe low-temperature phase transitions in certain regions, such as at $B_\perp = 2$ T. To address this, a smaller single crystal (0.05 mg) was selected for MCE testing, and the results are presented in **Figure 4i**. Because the ADR device inevitably absorbs and releases a small amount of heat, the magnetization process of the small crystal cannot be considered entirely adiabatic. Therefore, the point with the largest temperature change (d$T$/d$B$) was taken as the critical point and is marked with a black triangle. The results show that the positions of $B_{c1}$ and $B_{c3}$ are consistent with the specific heat phase diagram (**Figure 3d**), and a new quantum critical point $B_{c2}$ is discovered. This finding suggests that there may be different magnetic ground states (**V**) between $B_{c2}$ and $B_{c3}$.

The above results and analysis reveal the three successive phase transitions under zero field and a complex magnetic phase diagram with five phase regions and three quantum critical points for GdBO$_3$. The substantial magnetic entropy changes are significantly amplified by the strong quantum fluctuations of the Gd-based TL in ultra-low temperature regions. Furthermore, the intricate magnetic behaviors of GdBO$_3$ suggest a potential ADR strategy, which could be employed in a superfluid magnetic pump to drive superfluid helium to lower-temperature regions.

**Table 1.** Parameters of selected Gd-based adiabatic demagnetization refrigeration materials.

| Refrigerants | $T_c$ (K) | $B$ (T) | $-\Delta S_{max}$ | |
| --- | --- | --- | --- | --- |
| | | | J Kg$^{-1}$ K$^{-1}$ | mJ cm$^{-3}$ K$^{-1}$ |
| Gd$_3$Ga$_5$O$_{12}$ (GGG)[6] | 0.8 | 7 | 38.4 | 272 |
| GdF$_3$[15] | 1.25 | 7 | 71.6 | 506 |
| Gd(OH)$_3$[16] | 0.94 | 7 | 62.0 | 346 |
| GdPO$_4$[17] | 0.77 | 7 | 62.0 | 376 |
| Gd(OH)F$_2$[18] | 0.5 | 7 | 76.2 | 466 |
| Gd(OH)CO$_3$[19] | 0.7 | 7 | 66.4 | 355 |
| Gd(HCOO)$_3$[20] | 0.8 | 7 | 55.9 | 216 |
| KBaGd(BO$_3$)$_2$[21] | 0.26 | 5 | 29.0 | 142 |
| GdBO$_3$ (this work) | 1.77, 0.88, 0.52 | 5 | 33.8 | 213 |
| | | 7 | 49.7 | 314 |
| | | 9 | 57.4 | 363 |

## 3. Conclusion

In summary, we have shown millimeter-size single crystals growth of GdBO$_3$ through the high-temperature flux method and structural validation via the X-ray Diffraction and IR spectra analyses. Subsequently, magnetic susceptibility confirmed the existence of field-induced antiferromagnetic pairing at $B_\perp$ = 1.6 T, $T$ = 1.8 K. Additionally, the thermodynamic analysis revealed multi-level magnetic phase transitions ($T_{N1}$ = 0.52 K, $T_{N2}$ = 0.88 K, and $T_{N3}$ = 1.77 K) with $B_\perp$. The systematical specific heat and MCE analysis identify 5 distinct magnetic phase regions (**I** - **V**) and 3 clear quantum critical points ($B_{c1}$, $B_{c2}$, and $B_{c3}$) near 100 mK. In addition, we evaluated the MCE intensity in GdBO$_3$ and successfully cooled the sample temperature to 50 mK at the quantum critical point $B_{c3}$ based on a self-designed ADR device. We present an effective strategy for establishing the large magnetic entropy induced by quantum critical points in TL magnets for ADR applications, and our results highlight the significant potential of GdBO$_3$ as a magnetic cooling material in the milli-Kelvin region.

## 4. Method

***Single crystal growth***: High-purity Gd$_2$O$_3$, H$_3$BO$_3$, LiF, and NaF were weighed and evenly ground with a molar ratio of (1-2) : (7-9) : (8-12) : (3-7) in an agate mortar and placed in a platinum crucible. The crucible was transferred into the muffle furnace at 900 °C, then heated to 1050 °C at a rate of 150 °C/h to obtain a homogeneous high-temperature solution. During the cooling process, the solution was cooled to 700 °C at a 3 °C/h rate during the single crystal growth process. Finally, it was cooled to room temperature through a fast-cooling rate of 60 °C/h. The platinum crucible was immersed in a 5% dilute hydrochloric acid aqueous solution in a glass beaker and boiled for 2 hours to clean the flux on the crystal surfaces.

*P-XRD*: High-quality single crystals were selected under an optical microscope for thorough grinding. The obtained powder sample was then spread flat on a special glass slide and installed on the sample stage of the Rigaku SmartLab 9 KW instrument. The Bragg-Brentano geometric optical path was chosen for diffraction experiments. The light source used Cu K$\alpha$ (K$\alpha$1 = 1.54056 Å, K$\alpha$2 = 1.54439 Å) radiation. The program settings were as follows: the scanning range was 10°~100°, the step length was 0.02°, and the scanning speed was 1.5°/min. Finally, the experimental data were refined using GSAS II software.[22]

*SC-XRD*: SC-XRD experiments were conducted on a Bruker D8 VENTURE diffractometer equipped with a PHOTON III CPAD detector. The experiments were performed at 100 K in a temperature-controlled system utilizing nitrogen (N$_2$). X-rays ($\lambda$ = 0.71073 Å) were generated using a graphite monochromated Mo target. Background, polarization, and Lorentz factor corrections were subsequently applied using the APEX4 software, and multi-scan absorption correction was carried out with the SADABS package. The direct method of the ShelXT program was used for the structural solution, and the ShelXL least squares refinement package of Olex2 software was used for structural refinement.[23] Finally, possible higher symmetries were tracked using the ADDSYM algorithm in the PLATON program, and no missing symmetries were found.[24]

*Magnetic measurement*: The direct-current (DC) magnetic susceptibilities were performed on the Physical Property Measurement System (PPMS) and Magnetic Property Measurement System (MPMS3) developed by Quantum Design. The vibrating sample magnetometer (VSM) option of PPMS (PPMS) and the Superconducting Quantum Interference Devices (SQUID) VSM of the MPMS3 magnetic measurement system were utilized. Based on the well-shaped crystal, the magnetic field direction was aligned parallel to the triangular lattice layer ($B$ in-plane or $B_{//}$) and perpendicular to the triangular lattice layer ($B$ out-of-plane or $B_\perp$) for testing, as illustrated in **Figure S8**.

*Specific heat measurement*: Specific heat of GdBO$_3$ was performed on He3 (0.5 K - 5 K) and dilution fridge (DR) modules (> 0.1 K) in PPMS with varying $B_\perp$ applied. A 0.43 mg crystal of GdBO$_3$ was selected for He3, while a 0.0425 mg (after correction) sample was chosen for DR. The specific heat of LuBO$_3$ was used as a deduction for the phonon-specific heat.[10]

*Fourier transform infrared spectrometer*: GdBO$_3$ crystals were ground into powder using agate mortar, mixed with dry KBr at a mass ratio 1:100, and pressed into thin sheets using a tablet press. Infrared spectra of the sample were recorded in the 400–1400 cm$^{-1}$ range using an Excalibur 3100 Fourier transform infrared (FT-IR) spectrometer.

*MCE measurement*: The MCE measurements were performed on the Physical Property Measurement System (PPMS) with a homemade sample puck. A single crystal (10 mg) was mounted on a sapphire-plate platform supported by a fiberglass frame. A RuO$_2$ thermometer was affixed to the back of the platform to monitor the temperature. The sample puck was connected to the cold bath inside the high-vacuum chamber of the dilution refrigerator, while the sample was thermally isolated from the cold bath. From 0.1 K to 4 K, at each base temperature, the magnetic field of 7 T was swept to zero at a rate of 60 Oe/s. The temperature changes of the sample during the magnetic field sweep were recorded to determine the MCE.

**Acknowledgments**

The authors acknowledge the financial support from the National Natural Science Foundation of China (Grants No. 22205091, No. 12134020, No. 12374146, and No. 12104255), and the National Key Research and Development Program of China (Grant No. 2021YFA1400400). W. Lin and N. Zhao contributed equally to this work.

**Conflict of Interest**

The authors declare no conflict of interest.

**Data Availability Statement**

The data that support the findings of this study are available from the corresponding author upon reasonable request.

Supporting Information

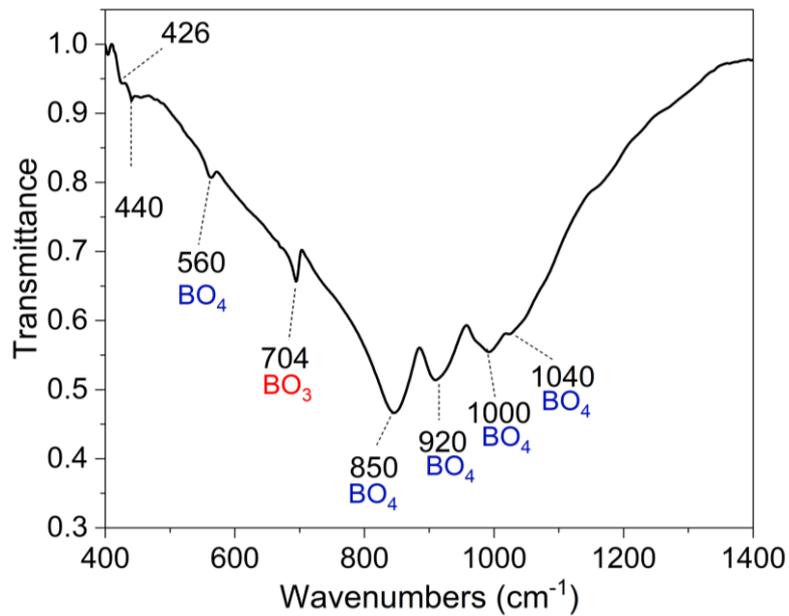

**Figure S1.** Fourier transform infrared of GdBO$_3$.

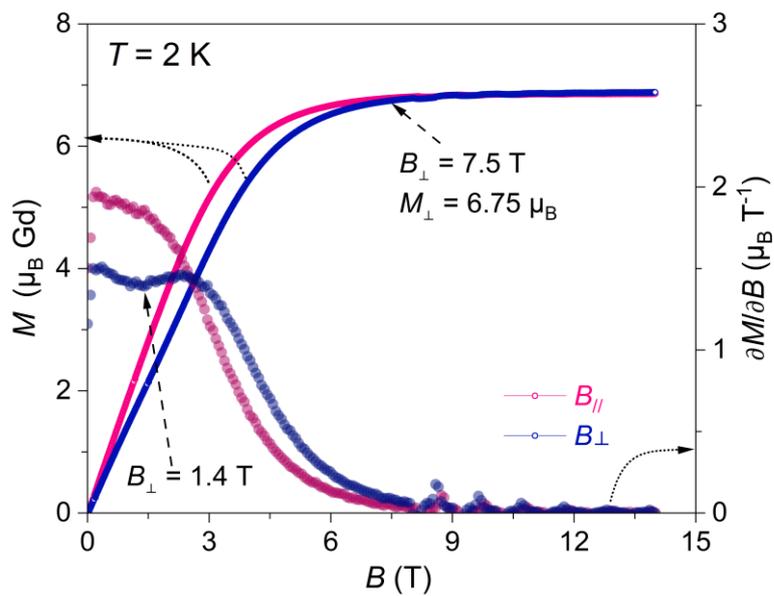

**Figure S2.** Field-dependent magnetization for $B_\parallel$ and $B_\perp$ at 2 K, measured by PPMS. The right *y*-axis presents the $\frac{\partial M}{\partial B}$ (corresponding to discrete dots).

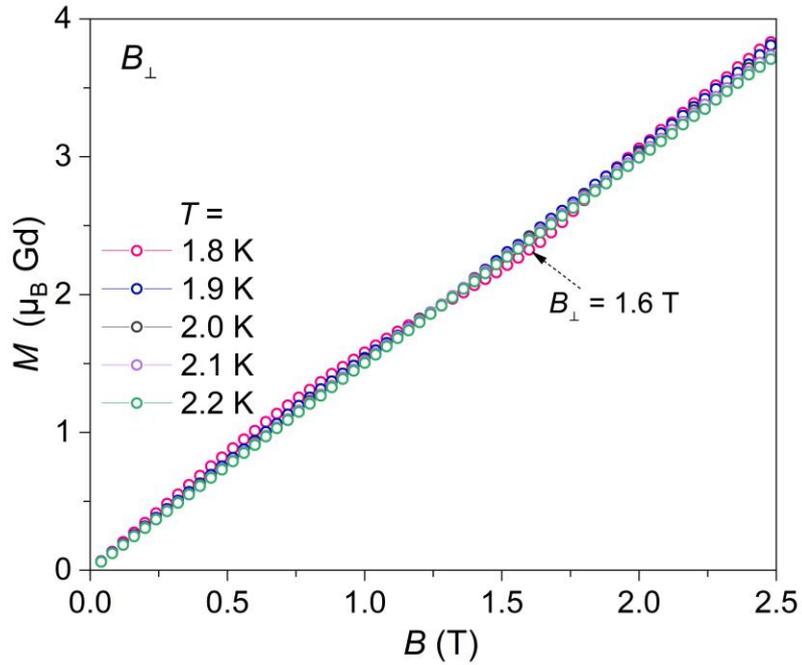

**Figure S3.** Magnetization of GdBO$_3$ with $B_\perp$.

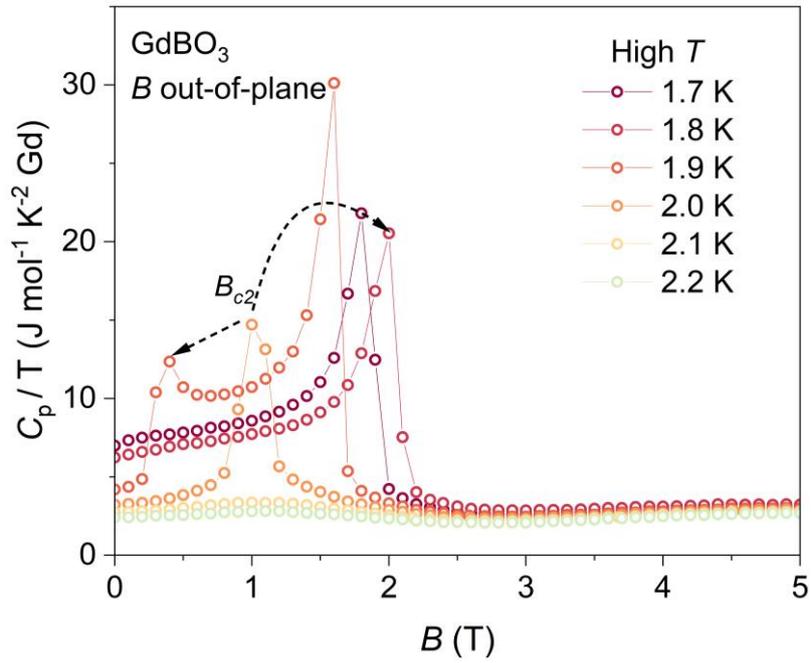

**Figure S4.** Field-dependence specific heat measured under 1.7 K – 2.2 K.

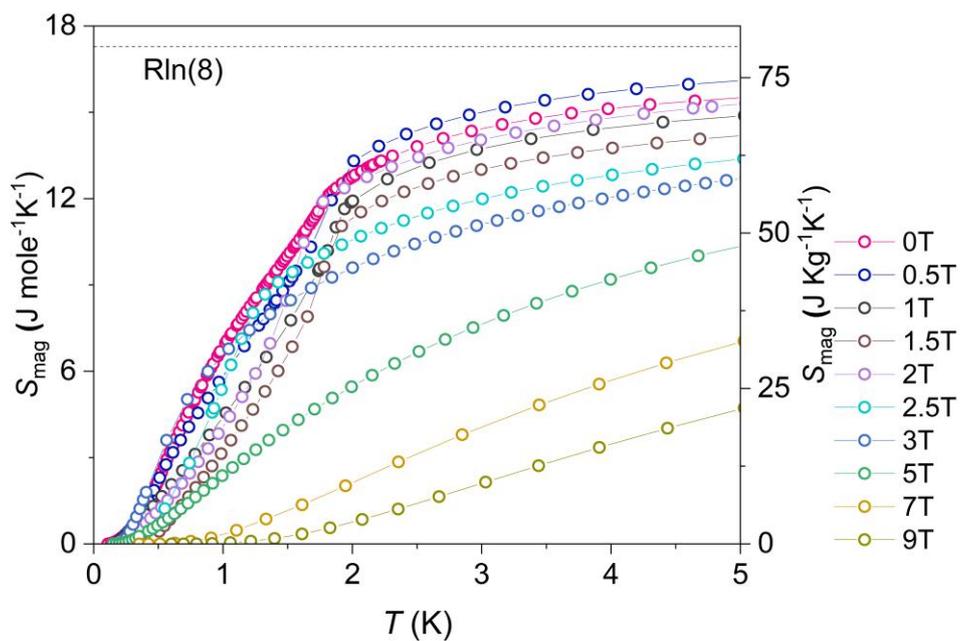

**Figure S5.** The magnetic entropy of GdBO$_3$.

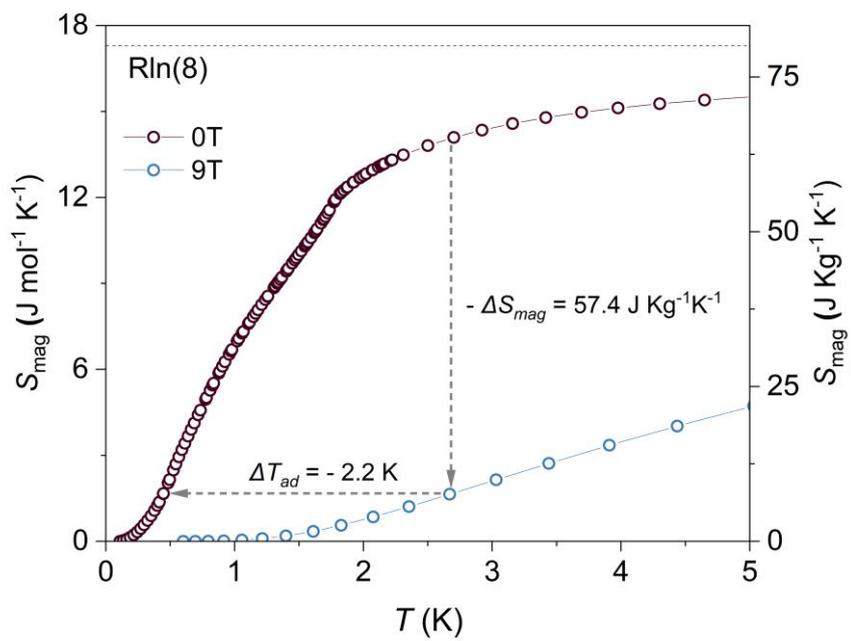

**Figure S6.** The principle of the magnetocaloric effect.

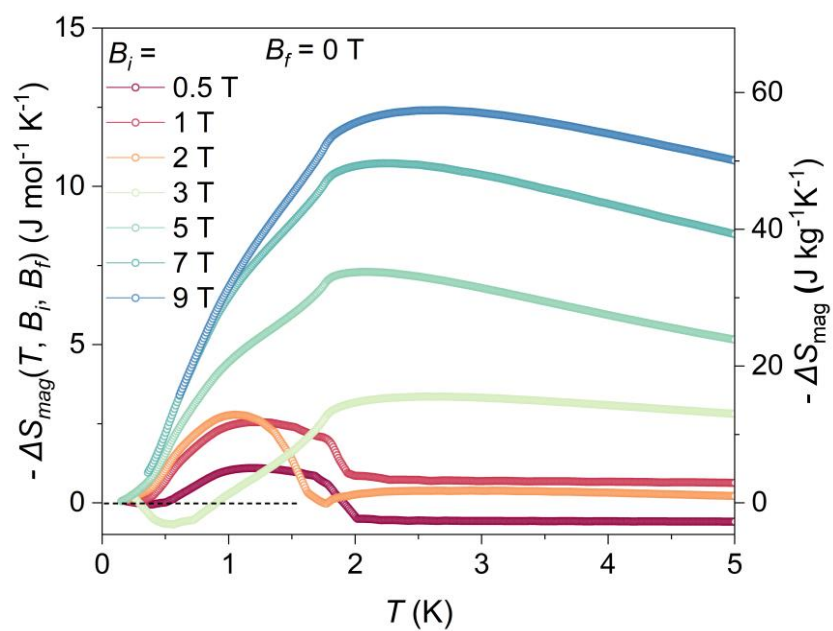

**Figure S7.** Magnetic entropy changes result in $\Delta S_{mag}$ vs. $T$ of GdBO$_3$ under final fields 0 T.

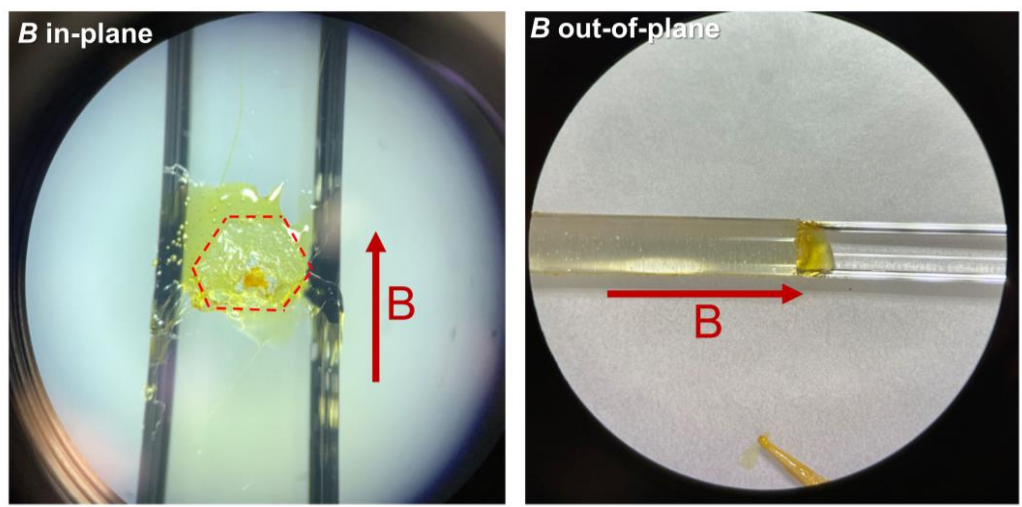

**Figure S8.** Details for measuring anisotropic magnetic susceptibilities of GdBO$_3$. *B* in-plane refers to the magnetic field parallel to the triangular lattice (TL) plane (001) and *B* out-of-plane direction refers to the magnetic field perpendicular to the TL plane.

**Table S1.** Crystal Data and Structure Refinements for GdBO$_3$.

| Empirical formula | GdBO$_3$ |
|---|---|
| Formula weight | 216.06 |
| Temperature/K | 100 |
| Crystal system | Monoclinic |
| Space group | *C2/c* |
| *a*/Å | 11.4684(1) |
| *b*/Å | 6.6318(8) |
| *c*/Å | 9.6703(1) |
| *β*/° | 112.842(4) |
| volume/Å$^3$ | 677.81(1) |
| Z | 4 |
| $\rho_{calc}$/g/cm$^3$ | 6.352 |
| $\mu$/mm$^{-1}$ | 29.042 |
| F(000) | 1116.0 |
| Mo K\α radiation/Å | 0.71073 |
| 2θ range for data collection/° | 7.26-61.04 |
| Goodness-of-fit on $F^2$ | 1.038 |
| Largest diff peak and hole (e/A$^3$) | 0.977 and -1.108 |
| Final *R* indexes [$I \geq 2\sigma(I)$] | R1=0.0260, wR2=0.0461 |
| Final *R* indexes [all data] | R1=0.0427, wR2=0.0528 |

**Table S2**. Fractional atomic coordinates and equivalent isotropic displacement parameters (Å$^2$) for GdBO$_3$.

| Atom | x | y | z | U$_{eq}$ |
|---|---|---|---|---|
| Gd1 | 0.250000 | 0.250000 | 0.000000 | 0.00314(10) |
| Gd2 | 0.08527(3) | 0.25548(4) | 0.49939(3) | 0.00335(9) |
| B1 | 0.1195(7) | 0.0358(10) | 0.2473(8) | 0.0073(13) |
| B2 | 0.000000 | 0.6768(15) | 0.250000 | 0.0062(18) |
| O1 | 0.1246(4) | 0.0894(7) | 0.1057(5) | 0.0045(8) |
| O2 | 0.2194(4) | 0.0906(7) | 0.3885(5) | 0.0052(8) |
| O3 | 0.0472(4) | 0.5705(7) | 0.3901(5) | 0.0048(8) |
| O4 | 0.3934(4) | 0.3104(7) | 0.2514(5) | 0.0079(9) |
| O5 | 0.000000 | 0.1312(9) | 0.250000 | 0.0057(12) |